\documentclass[12pt,twocolumn]{iopart}
\usepackage{graphicx}
\usepackage{epsfig}
\usepackage{iopams}
\usepackage{subfigure}
\usepackage{multicol} 
\usepackage{cite}

\newcommand{\kommentar}[1]{}

\newcommand{\ket}[1]{\ensuremath{|#1\rangle}}

\newcommand{\ketbra}[1]{\ensuremath{| #1 \rangle \langle #1 |}}

\newcommand{\eins}{\ensuremath{1\!\!1}}

\newcommand{\WW}{\ensuremath{\mathcal{W}}}

\newcommand{\BE}{\begin{equation}}
\newcommand{\EE}{\end{equation}}
\newcommand{\be}{\begin{equation}}
\newcommand{\ee}{\end{equation}}
\newcommand{\bea}{\begin{eqnarray}}
\newcommand{\eea}{\end{eqnarray}}
\newcommand{\bean}{\begin{eqnarray*}}
\newcommand{\eean}{\end{eqnarray*}}

\newcommand{\bc}{\begin{center}}
\newcommand{\ec}{\end{center}}

\newcommand{\vr}{\ensuremath{\varrho}}

\begin{document}

\title[Robustness of multiparticle entanglement]{Robustness 
of multiparticle entanglement: specific entanglement classes 
and random states}

\author{Mazhar Ali\footnote{Permanent address: Department of Electrical Engineering, COMSATS Institute of Information Technology, 22060 Abbottabad, Pakistan.} and Otfried G\"uhne}
\address{Naturwissenschaftlich-Technische Fakult\"at, Universit\"{a}t Siegen, Walter-Flex-Stra\ss e 3, 57068 Siegen, Germany} 

\begin{abstract}
We investigate the robustness of genuine multiparticle 
entanglement under decoherence. We consider different 
kinds of entangled three- and four-qubit states as well 
as random pure states. For amplitude damping noise, 
we find that the W-type states are most robust, while other 
states are not more robust than generic states. For phase 
damping noise the GHZ state is the most robust state, and 
for depolarizing noise several states are significantly more
robust than random states. 
\end{abstract}


\section{Introduction}

Entanglement between different particles is one of the peculiar 
features of quantum physics and its characterization is an active 
field of research \cite{Horodecki-RMP-2009, gtreview}. In theory, pure state 
entanglement can be used for tasks like quantum teleportation or 
cryptography, but in real implementations one cannot avoid interactions 
with the environment, leading to noise and decoherence. Therefore it is
important to study the robustness of entangled states and many research
efforts have been undertaken in this direction \cite{lifetime, acindec, 
bipartitedec,lowerbounds, Guehne-PRA78-2008, furtherdec}.

Several problems concerning the robustness of entanglement under decoherence 
have been discussed so far. 
Many works considered the life time of entanglement under decoherence 
\cite{lifetime}, but it should be noted that the life time may not 
characterize the decoherence process well, since it can happen that 
the life time of entanglement is large, but the actual amount of 
entanglement is small already after a short time, making the decohered
state practically useless for information processing tasks \cite{acindec}. 
Here, the life time of entanglement denotes the time with a nonzero value 
of the chosen measure of entanglement.
Further works considered bipartite aspects of the entanglement of several 
particles \cite{bipartitedec}, but this can only 
give a partial characterization, since multiparticle entanglement is known 
to be different from entanglement between all bipartitions \cite{gtreview}.
A further central problem behind existing studies is that the theory of 
multiparticle entanglement is still not fully developed, so for many 
cases one can only make statements about lower bounds on entanglement, 
but not the actual value \cite{lowerbounds}. The exact calculation of a 
multiparticle entanglement measure was, so far, only possible for special 
states and decoherence models \cite{Guehne-PRA78-2008}. Finally, in order 
to make a fair statement about the robustness of a specific state one has 
to compare it with random states. Due to the difficulty in evaluating 
multiparticle entanglement measures, this has not been investigated so 
far. 

In this note we study the robustness of various prominent multi-qubit
states as well as random states under local decoherence. Our work is
enabled by recent progress in the theory of multiparticle entanglement,
especially the computable entanglement monotone for genuine multiparticle
entanglement from Ref. \cite{Bastian-PRL106-2011}. 
To calculate the robustness of this monotone, we have chosen all those 
time scales for which the values of monotone are not vanishingly small.   
We study different models of decoherence and identify the most robust states for this scenarios.
For amplitude damping noise, we find that the W-type states are most robust, 
for phase  damping noise the GHZ state is the most robust state, and 
for depolarizing noise several states are significantly more
robust than random states. 

The reader should be aware of the fact that the term "robustness of 
entanglement" was used in Ref.~\cite{Vidal-PRA59-1999} as a kind of 
quantification of entanglement, but our approach is different from 
this work, since we consider the behaviour under decoherence. Also
in this paper we speak of multiparticle systems and multiparticle 
entanglement, but we essentially discuss two-level quantum systems 
(or qubits). This means that our results may also be applied to 
multi-qubit systems, where the qubits are not separated as particles. 

This paper is organized as follows. In Section 2.1, we describe 
our physical model to obtain the dynamics 
of an arbitrary density matrix. In Section 2.2, we briefly review the 
concept of multiparticle entanglement and also review the derivation 
of entanglement monotone. In Sections 2.3 - 2.5, we define the investigated
states including the generation method of random pure states and weighted graph states. We present the main results in Sec. \ref{Sec:Results}. 
Finally, we offer some conclusions in Sec. \ref{Conc}.

\section{Preliminaries}

\subsection{Local decoherence models for multiparticle states} 
\label{Sec:Model}

We consider $N$ qubits (e.g., $N$ two level atoms) which 
are coupled to their own local reservoirs. The reservoirs are 
assumed to be independent from each other. We assume weak coupling 
between each qubit and the corresponding reservoir and no back action 
effect of the qubits on the reservoirs. We also assume that the 
correlation time between the qubits and the reservoirs is much shorter 
than the characteristic time of the evolution so that the Markovian 
approximation is valid. The interactions of the physical system with 
environment can be studied via various techniques, for example, 
solving a master equation, the Kraus operator formalism, or quantum trajectories, 
etc. We work in the Kraus  operator formalism. The time evolution 
of an initial density matrix can be written as
\begin{equation}
\varrho(t) = \sum_i  \, K_i(t) \, \varrho(0) \, K_i^\dagger(t), 
\label{Eq:TE}
\end{equation}
where $K_i(t)$ are the Kraus operators, satisfying the 
normalization condition $\sum_i \, K_i^\dagger(t) \, K_i(t) =  \eins$. 
The precise form of these Kraus operators are given as 
$
K_i (t) = \omega_{i_1}^A  \otimes \omega_{i_2}^B \otimes \cdots \otimes \omega_{i_N}^N,
$
where $\omega_i^j$ are the single-qubit Kraus operators acting on the 
$j$th qubit. 

In the following, we consider three models. For {\it amplitude damping},
there are two Kraus operators for a single qubit,
\be
\omega_1^j = \left( 
\begin{array}{cc}
1 & 0 \\ 
0 & \gamma_j
\end{array}
\right) \, , \quad \omega_2^j = \left( 
\begin{array}{cc}
0 & \sqrt{1-\gamma_j^2} \\ 
0 & 0
\end{array}
\right), 
\ee
where $\gamma_j = {\rm e}^{- \Gamma_j t/2}$. For {\it phase damping},
the corresponding single-qubit Kraus operators are given as 
\be
\omega_1^j = \left( 
\begin{array}{cc}
1 & 0 \\ 
0 & \gamma_j
\end{array}
\right) \, , \quad \omega_2^j = \left( 
\begin{array}{cc}
0 & 0 \\
0 & \sqrt{1-\gamma_j^2} 
\end{array}
\right). 
\ee
Finally, for the {\it depolarizing channel} 
there are four single-qubit Kraus operators, 
\begin{eqnarray}
\omega_1^j = \sqrt{1-q} \, {\eins},\, \omega_2^j = \sqrt{\frac{q}{3}} \, \sigma_x ,\, \omega_3^j = \sqrt{\frac{q}{3}} \, \sigma_y ,\,\omega_4^j = \sqrt{\frac{q}{3}} \, \sigma_z , 
\end{eqnarray}
where $q = 3 p/4$ with $p = 1-\gamma_j$, and $\sigma_x$, $\sigma_y$, $\sigma_z$ are the Pauli matrices. Note that in ion-trap experiments, 
amplitude damping is the typical noise, while for photonic experiments
phase damping is more relevant. For the case of $N$-qubit states, 
there are $2^N$  global Kraus operators $K_i(t)$ for the amplitude 
and phase damping channels and $4^N$ Kraus operators for the 
depolarizing channel. For the sake of simplicity we assume onwards that 
$\gamma_A = \gamma_B = \cdots = \gamma_N = \gamma$.

The time evolved density matrix for a single qubit can directly 
be computed. Under amplitude damping it is given as 
\be
\varrho (t) = \left( \begin{array}{cc}
\varrho_{11} + \varrho_{22} (1-e^{- \gamma t}) & \varrho_{12} \, e^{- \gamma t/2 } \\ 
\varrho_{21} \, e^{- \gamma t/2 } & \varrho_{22} \, e^{- \gamma t} 
\end{array} \right),
\ee
whereas the time evolved density matrix for a single qubit state under phase damping is given as  
\be
\varrho (t) = \left( \begin{array}{cc}
\varrho_{11} & \varrho_{12} \, e^{- \gamma t/2 } \\ 
\varrho_{21} \, e^{- \gamma t/2 } & \varrho_{22}  
\end{array} \right). 
\ee
Finally, the density matrix for a single qubit state under 
depolarizing noise  is 
\be
\varrho (t) 
= \left( \begin{array}{cc}
\varrho_{11} + p (\frac{1}{2} - \varrho_{11}) & \varrho_{12} - p \, \varrho_{12} \\ 
\varrho_{21} - p \, \varrho_{21} & \varrho_{22} + p (\frac{1}{2} - \varrho_{22})  
\end{array} \right)  = (1-p) \varrho + p \frac{\eins}{2}. 
\ee
For more qubits, the calculation of density matrices is straightforward.

\subsection{Genuine multiparticle entanglement and 
the multiparticle negativity} 
\label{Sec:GME}

Let us first recall the basic definitions for genuine multiparticle 
entanglement. We explain the main ideas by considering three 
particles $A$, $B$, and $C$,  the generalization
to more parties is straightforward. A state is separable with respect 
to some bipartition, say, $A|BC$, if it is a mixture of product 
states with respect 
to this partition, that is, $\varrho = \sum_k \, q_k \, |\phi_A^k \rangle\langle \phi_A^k| \otimes |\psi_{BC}^k \rangle\langle \phi_{BC}^k|$, where the $q_k$ form 
a probability distribution. We denote these states as $\varrho_{A|BC}^{sep}$. 
Similarly, we can define separable states for the two other  bipartitions $\varrho_{B|AC}^{sep}$ and $\varrho_{C|AB}^{sep}$. Then a state is called biseparable 
if it can be written as a mixture of states which are separable with respect 
to different bipartitions, that is 
\be
 \varrho^{bs} = p_1 \, \varrho_{A|BC}^{sep} + p_2 \, \varrho_{B|AC}^{sep} + p_3 \, \varrho_{C|AB}^{sep}\,.
\ee
Finally, a state is called genuinely multiparticle entangled 
if it is not biseparable. In remainder of this paper, we always mean genuine
multiparticle entanglement when we talk about entanglement. 

Recently, a powerful technique has been worked out to detect and 
characterize multiparticle entanglement \cite{Bastian-PRL106-2011}. 
The method is to use positive partial transpose mixtures (PPT mixtures). 
Recall that a two-party state $\varrho = \sum_{ijkl} \, \varrho_{ij,kl} \, |i\rangle\langle j| \otimes |k\rangle\langle l|$ is PPT if its partially transposed matrix 
$\varrho^{T_A} = \sum_{ijkl} \, \varrho_{ji,kl} \, |i\rangle\langle j| \otimes |k\rangle\langle l|$ has no negative eigenvalues. A well known fact is 
that separable states are always PPT \cite{peresppt}. The 
set of separable states with respect to some partition is therefore 
contained in a larger set of states which has a positive partial 
transpose for that bipartition. 

We denote the states which are PPT with respect to fixed bipartition by 
$\varrho_{A|BC}^{ppt}$, $\varrho_{B|AC}^{ppt}$, and $\varrho_{C|AB}^{ppt}$ and 
ask the question whether a state can be written as
\begin{eqnarray}
\varrho^{pptmix} = p_1 \, \varrho_{A|BC}^{ppt} + p_2 \, \varrho_{B|AC}^{ppt} + p_3 \, \varrho_{C|AB}^{ppt}\,.
\end{eqnarray}
Such a mixing of PPT states is called a PPT mixture. 
The genuine multiparticle entanglement of four or more particles 
can be detected and quantified in an analoguous manner by considering 
all bipartitions (like one particle vs. $N-1$ particles, 
two particles vs. $N-2$ particles, etc.).

Obviously, any biseparable state is a PPT mixture, therefore any state which is not 
a PPT mixture is guaranteed to be genuinely multiparticle entangled. 
The major advantage of considering PPT mixtures instead of biseparable 
states comes from the fact that PPT mixtures can be fully characterized with 
the method of semidefinite programming (SDP), a standard  method
in convex optimization \cite{sdp}. In general, the set of PPT mixtures is 
a very good approximation to the set of biseparable states and delivers 
the best known separability criteria for many cases, but
it must be stressed that there are multiparticle entangled states which 
are PPT mixtures \cite{Bastian-PRL106-2011}.

Let us briefly describe the SDP. As shown in Ref.~\cite{Bastian-PRL106-2011}, 
a state is a PPT mixture iff the following optimization problem 
\begin{eqnarray}
 \min {\rm Tr} (\WW \varrho)
\end{eqnarray}
under the constraint that for all bipartitions $M|\bar{M}$
\begin{eqnarray}
 \WW = P_M + Q_M^{T_M},
 \quad \mbox{ with }
 0 \leq P_M\,\leq \eins \mbox{ and }
 0 \leq  Q_M  \leq \eins\, 
\end{eqnarray}
has a positive solution. The constraints just state that
the considered operator $\WW$ is a decomposable entanglement
witness for any bipartition. If this minimum is negative then 
$\varrho$ is not a PPT mixture and hence genuinely multiparticle 
entangled. Since this is a semidefinite program, the minimum can 
be efficiently computed and the optimality of the solution can
be certified \cite{sdp}. For solving the SDP we used the programs YALMIP
and SDPT3 \cite{yalmip}, a ready-to-use implementation is freely 
available \cite{pptmix}.

For us, it is important that this approach can be used to {\it quantify} 
genuine entanglement. In fact the absolute value of the above 
minimization was shown to be an entanglement monotone for genuine multiparticle
entanglement \cite{Bastian-PRL106-2011}. In the following, we will denote 
this measure by $E(\varrho)$.
For bipartite systems, this monotone is equivalent to the so-called
{negativity} \cite{Vidal-PRA65-2002}. For a system of 
qubits, this measure is bounded by  $E(\varrho) \leq 1/2$
\cite{bastiangraph}.

\subsection{Investigated quantum states}
\label{Sec:RIQS}

Let us introduce the multiparticle entangled states 
which we  study in this paper. Two important types of states
are the GHZ states and the W states for $N$ qubits, 
\begin{eqnarray}
|GHZ_N \rangle &=& \frac{1}{\sqrt{2}}(\ket{00...0} + \ket{11...1}),
\nonumber
\\
\ket{W_N}&=&\frac{1}{\sqrt{N}}(\ket{00...001} + \ket{00...010} +\ket{10...000}).
\label{Eq:GHZ3Qb1}
\end{eqnarray}
For the GHZ state, the entanglement monotone has a value of
$E(\ketbra{GHZ_N}) = 1/2$, while for the W state, its 
value is
$E(\ketbra{W_3}) \approx 0.443$ and $E(\ketbra{W_4}) \approx 0.366$.

For the case of four qubits, several other states are interesting and 
have been discussed in the literature. These states are the 
Dicke state
$\ket{D_{2,4}}$, the four-qubit singlet state $|\Psi_{S,4}\rangle$,
the cluster state $\ket{CL}$ and the so-called $\chi$-state 
$\ket{\chi_4}$. They are explicitly given by:
\begin{eqnarray}
|D_{2,4} \rangle 
&=&\frac{1}{\sqrt{6}} [ |0011 \rangle + |1100\rangle + |0101 \rangle + |0110\rangle + |1001 \rangle + | 1010\rangle] \,,
\nonumber
\\
|\Psi_{S,4}\rangle &=& \frac{1}{\sqrt{3}} [ |0011 \rangle + |1100\rangle - \frac{1}{2} ( \, |0101 \rangle + |0110\rangle  + |1001 \rangle + |1010\rangle)] \,,
\nonumber
\\
|CL \rangle &=& \frac{1}{2}
[|0000 \rangle + |0011\rangle + |1100 \rangle - |1111\rangle],
\nonumber
\\
|\chi_4 \rangle &=& \frac{1}{\sqrt{6}} 
[\sqrt{2} | 1111 \rangle + |0001 \rangle + |0010\rangle + |0100 \rangle + |1000\rangle]. 
\end{eqnarray}
Note that all of these states have the maximum value of entanglement
$E (\ketbra{D_{2,4}}) = E (\ketbra{\Psi_{S,4}}) = E (\ketbra{CL}) 
= E (\ketbra{\chi_{4}}) = 1/2$. Further entanglement properties of 
these states are reviewed in Ref.~\cite{gtreview}.

\subsection{Weighted graph states}

Weighted graph states form a family of multi-qubit states that 
includes states with a large variety of entanglement features 
(such as GHZ and cluster states) \cite{Hartmann-JPB40-2007, Hein-arXiv2006}. 
In the past, the structure of weighted graph states made it possible 
to deal with thousands of spins (spin gases) to study their entanglement 
features \cite{Hartmann-JPB40-2007}. 

To define weighted graph states, consider a graph as a set of vertices and edges. 
Physically, the vertices denote the physical systems (qubits), whereas the edges 
represent the interactions among physical systems. In the beginning, one prepares 
all the qubits in the state $\ket{+}=(\ket{0}+\ket{1})/\sqrt{2}.$ Then, if 
two qubits $k,l$ are connected with an edge, one applies an interaction 
according to the Hamiltonian
\begin{eqnarray}
H_{kl}  = 
\frac{1}{4} \, (\eins - \sigma_z^{(k)}) \otimes (\eins - \sigma_z^{(l)}) \, ,
\end{eqnarray}
This leads to a unitary transformation of the type $U_{kl} = {\rm e}^{−i \, \phi_{kl} \, H_{kl}}$, where $\phi_{kl}$ is the interaction time. The resulting state is then called
the weighted graph state, and it can be expressed as
\begin{eqnarray}
|G\rangle = \bigotimes_{k,l} U_{kl} (\phi_{kl}) \, |+\rangle^{\otimes N} \, ,
\end{eqnarray}
In this way, the weighted graph state is uniquely determined by the $N(N-1)/2$
parameters $\phi_{kl}$. Clearly, weighted graph states  form only a small subset 
of all pure states (which are described by $2^N-1$ parameters), but many interesting
states fall in this class. For generating random weighted graph states, we have chosen the interaction times $\phi_{kl} \in [0, 2 \pi]$ uniformly distributed in the interval. 

Three more remarks are in order. First, if one considers only the possibilities
$\phi_{kl} = \pi$ or $\phi_{kl} = 0$, the usual graph states (to which the GHZ
and cluster state belong) emerge. Second, it should be noted that the unitaries
$U_{kl}$ commute, so the temporal order of the interaction does not matter. Finally, 
some generalizations of weighted graph states have been proposed and investigated
recently \cite{LME}.

\subsection{Random pure states}

Finally, let us describe how we have generated random pure states. 
A state vector randomly distributed according to the Haar measure  
can be generated in the following way \cite{Toth-Arxiv}: First, one generates 
a vector such that both the real and the imaginary parts of the vector 
elements are Gaussian distributed random numbers  with a zero mean and unit 
variance. Second we normalize the vector. It is easy to prove that the random 
vectors obtained this way are equally distributed on the unit sphere 
\cite{Toth-Arxiv}. Note that this generates random pure states in the 
global Hilbert space of three- and four qubits, so the unit sphere is 
not the Bloch ball.

\section{Results} 
\label{Sec:Results}

In this section we study the robustness of entanglement of three and four qubits 
under local decoherence. Before studying the different states and models, let us 
define how one can quantify the robustness. First, as already mentioned, the 
life-time of entanglement may lead to inconclusive results, since state under 
decoherence may be entangled for a long time, but the amount of entanglement 
may be nearly zero and therefore of little use for quantum information processing 
tasks. Second, the lifetime of entanglement is clearly not a reasonable figure
of merit, if the initial states that should be compared have already a different
amount of entanglement. For the same reason, also the comparison of the values of
$E[\varrho_i(t)]$ for different $t$ and initial states $i$ is not useful.

\begin{figure}[t!]
\scalebox{2.0}{\includegraphics[width=1.95in]{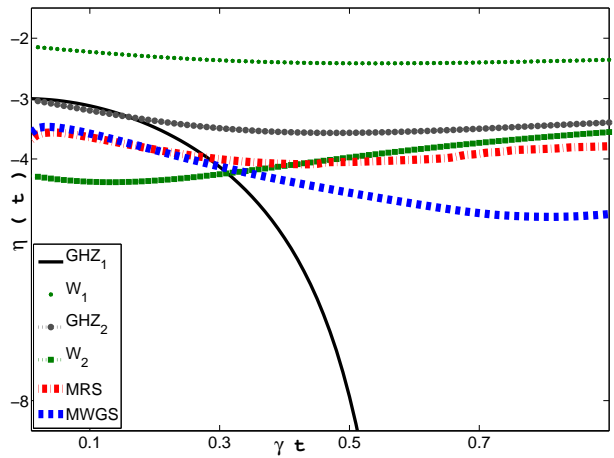}}
\centering
\caption{Logarithmic derivative of entanglement monotone for various three-qubit 
states under amplitude damping, plotted against parameter $\gamma t$. See text
for further details}
\label{Fig:R3QbAD1}
\end{figure}

In our approach, we consider the logarithmic derivative 
\begin{eqnarray}
\eta(t) = \frac{d\,(\ln[E(t)])}{dt} = \frac{d/dt \,[E(t)]}{E(t)} \, , 
\end{eqnarray}
where $E(t)$ is the entanglement monotone \cite{Guehne-PRA78-2008}. 
This describes the relative decay of entanglement, and allows to compare states with a different initial entanglement. In this way, the introduction of the logarithmic derivative makes sure that comparative change in monotone is meaningful for several different quantum states.
Note that for a state where the entanglement just decays exponentially, $\eta(t)$
is constant and the inverse of the half-life.

\subsection{Amplitude damping}\label{SSec:AD} 

\subsubsection{Three qubits}
Let us first consider the effects of amplitude damping on quantum 
states of three qubits. Fig.~\ref{Fig:R3QbAD1} shows the logarithmic derivative 
$\eta (t)$ of the entanglement monotone,  plotted against the 
parameter $\gamma t$ for different types of states. In this figure 
we show this parameter for the usual GHZ and W state (denoted by $GHZ_1$ 
and $W_1$). In addition, we computed the value for two different forms of
the GHZ and W state, namely $|GHZ_2\rangle = 1/\sqrt{2} (|001 \rangle + |110\rangle)$
and $|W_2\rangle = 1/\sqrt{3} (|110 \rangle + |101\rangle + |011 \rangle).$
Finally, we also give the mean value for random pure states (MRS) and 
random weighted graph states (MWGS). For computing these mean values, 
we considered 100 realizations of the respective states, an explicit 
figure is given in the Appendix. From these data we also obtain an 
error estimate to indicate the reliability of measure. 
This can, for instance, be defined as a confidence interval
\begin{eqnarray}
 CI = \mu \, \pm \, \sqrt{\delta} \, ,
 \label{interval}
\end{eqnarray}
where $\mu$ stands for mean value and $\delta$ for variance of 
quantity being measured. Note, however, that this is not a confidence
interval in the mathematical sense.

It should be mentioned that if the entanglement vanishes, the 
quantity $\eta(t)$ diverges. In our investigations we always 
consider timescales which are much shorter than the life time of
entanglement, so this effect does not occur in our figures. Only
for random states, we found singular cases (with a probability less 
than 1\%) where the entanglement vanishes fast, these are then 
not included in our sample of random states.

\begin{figure}[t!]
\centering
\subfigure[]{\includegraphics[width=7.7cm]{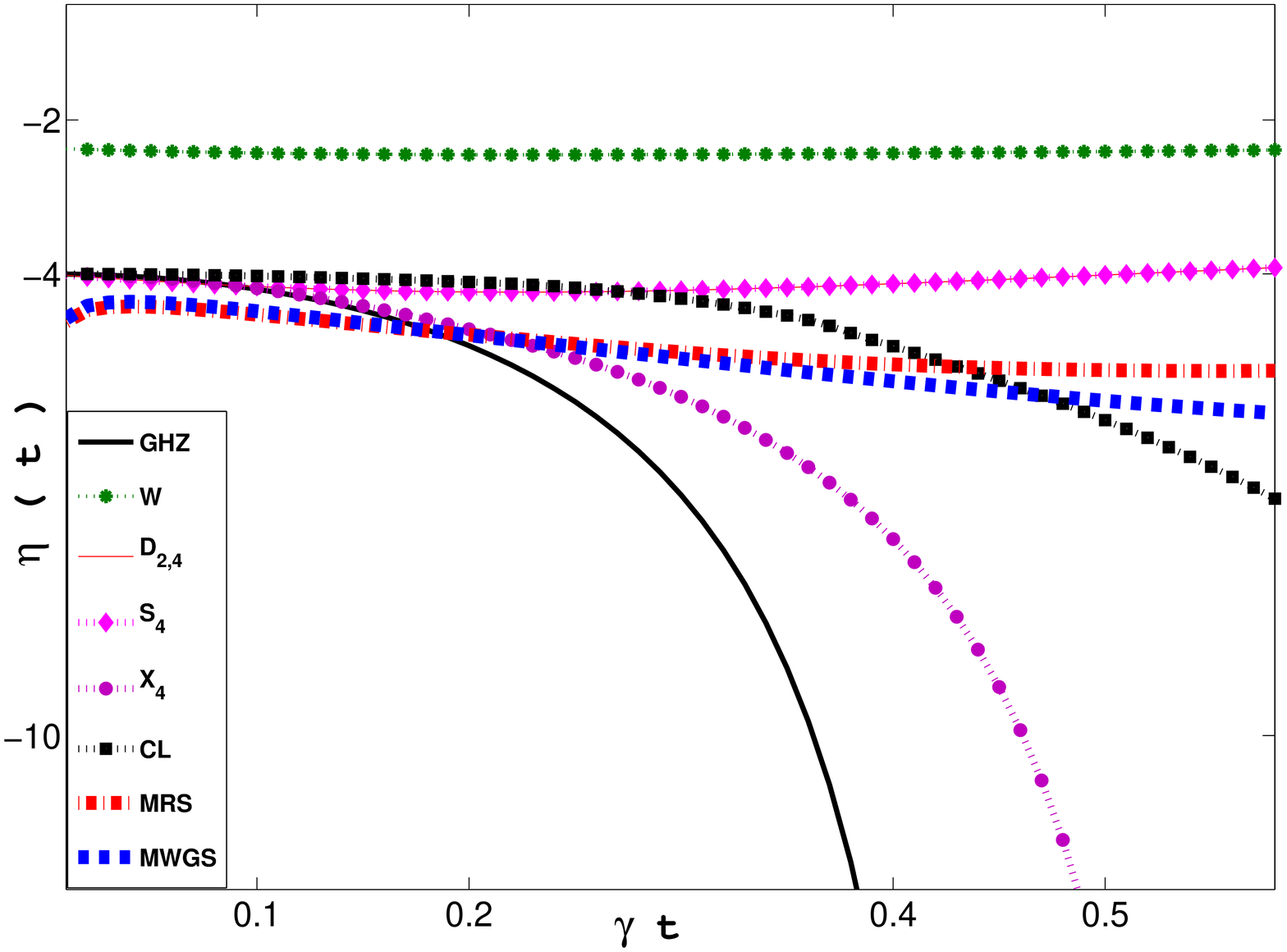}}
\subfigure[]{\includegraphics[width=7.7cm, height=5.3cm]{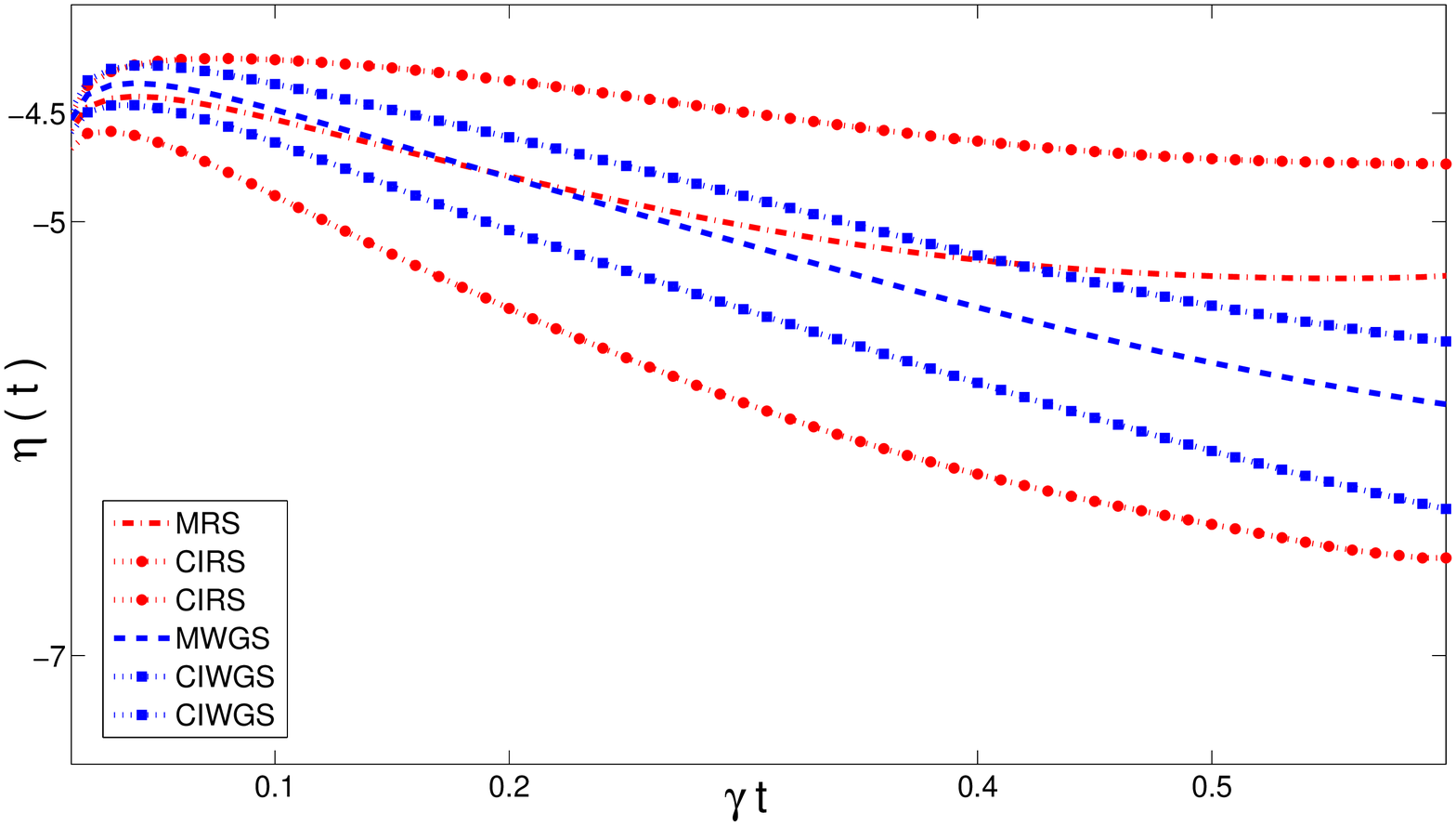}}
\caption{(a) Logarithmic derivative $\eta (t)$ for several four-qubit states 
under amplitude damping against parameter $\gamma t$. (b) Mean values and confidence 
intervals of $\eta (t)$ for random pure states and random weighted-graph states.
See text for further details.}
\label{Fig:R4QbAD1}
\end{figure}

From Fig.~\ref{Fig:R3QbAD1} and Fig.~\ref{Fig:rand} in the Appendix  
we can conclude two things: First, the W state is
for this type of decoherence clearly the most robust state. The other states 
do not deviate significantly from random states. Second, for nearly all states
the decay of entanglement is roughly exponential, while for the GHZ state it is 
super-exponential.

\subsubsection{Four qubits}
For the case of four qubits, the main results are given in Fig.~\ref{Fig:R4QbAD1}.
Here, we consider all the four-qubit states as introduced in Section 2.3: 
The W state, the GHZ state, the Dicke state $D_{2,4}$, the singlet state 
$S_4$, the $\chi_4$-state $X_4$ and the cluster state CL. Again, we compare 
them with random pure states and 
random weighted graph states. The conclusions are similar as for the three-qubit case: 
The entanglement present in the  W state is most robust against decoherence, the other 
states are, in this respect, not significantly different from random states. 

\subsection{Phase damping}
\label{SSec:PD}

\begin{figure}[t!]
\centering
\subfigure[]{\includegraphics[width=7.7cm]{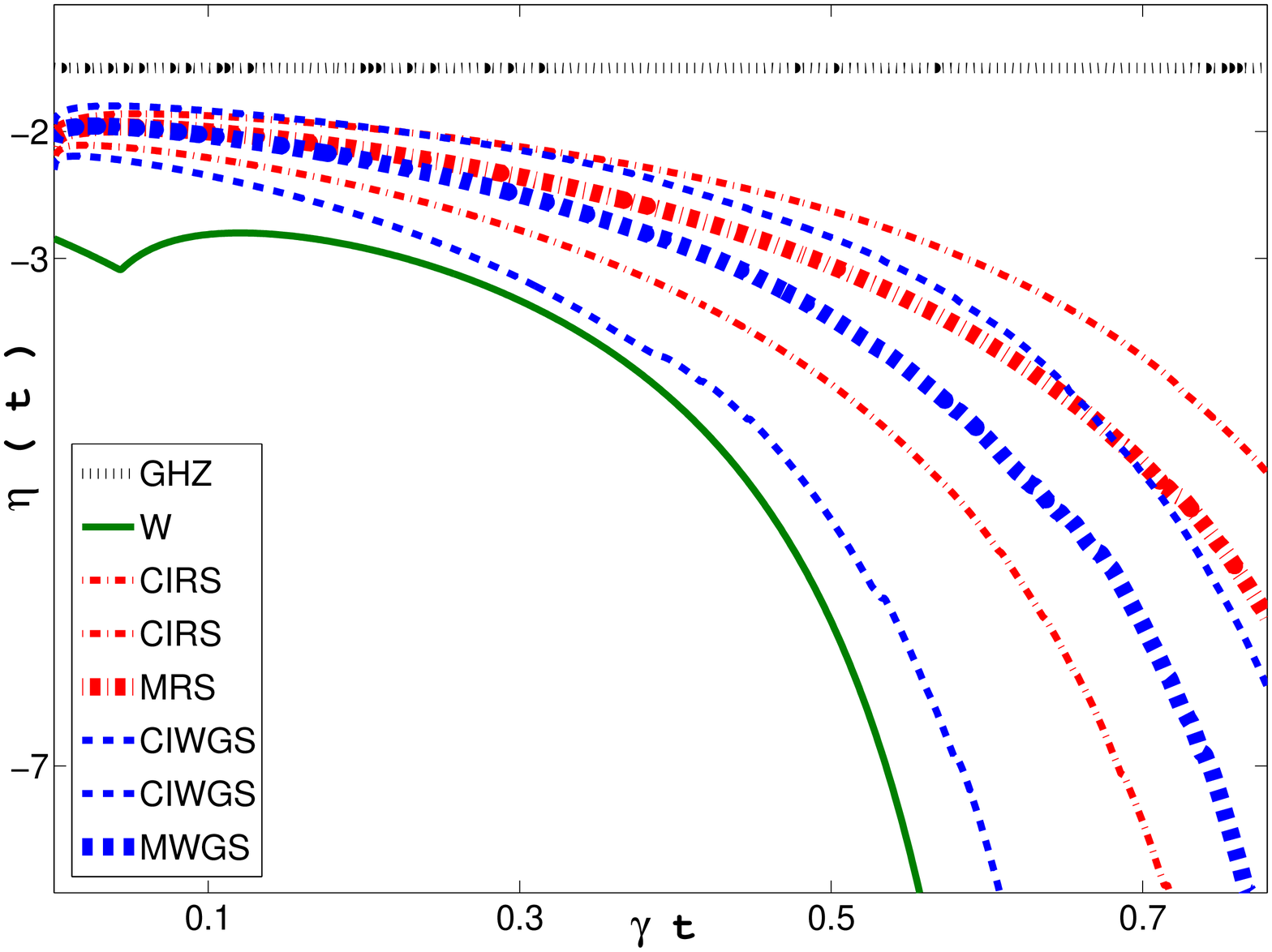}}
\subfigure[]{\includegraphics[width=7.7cm]{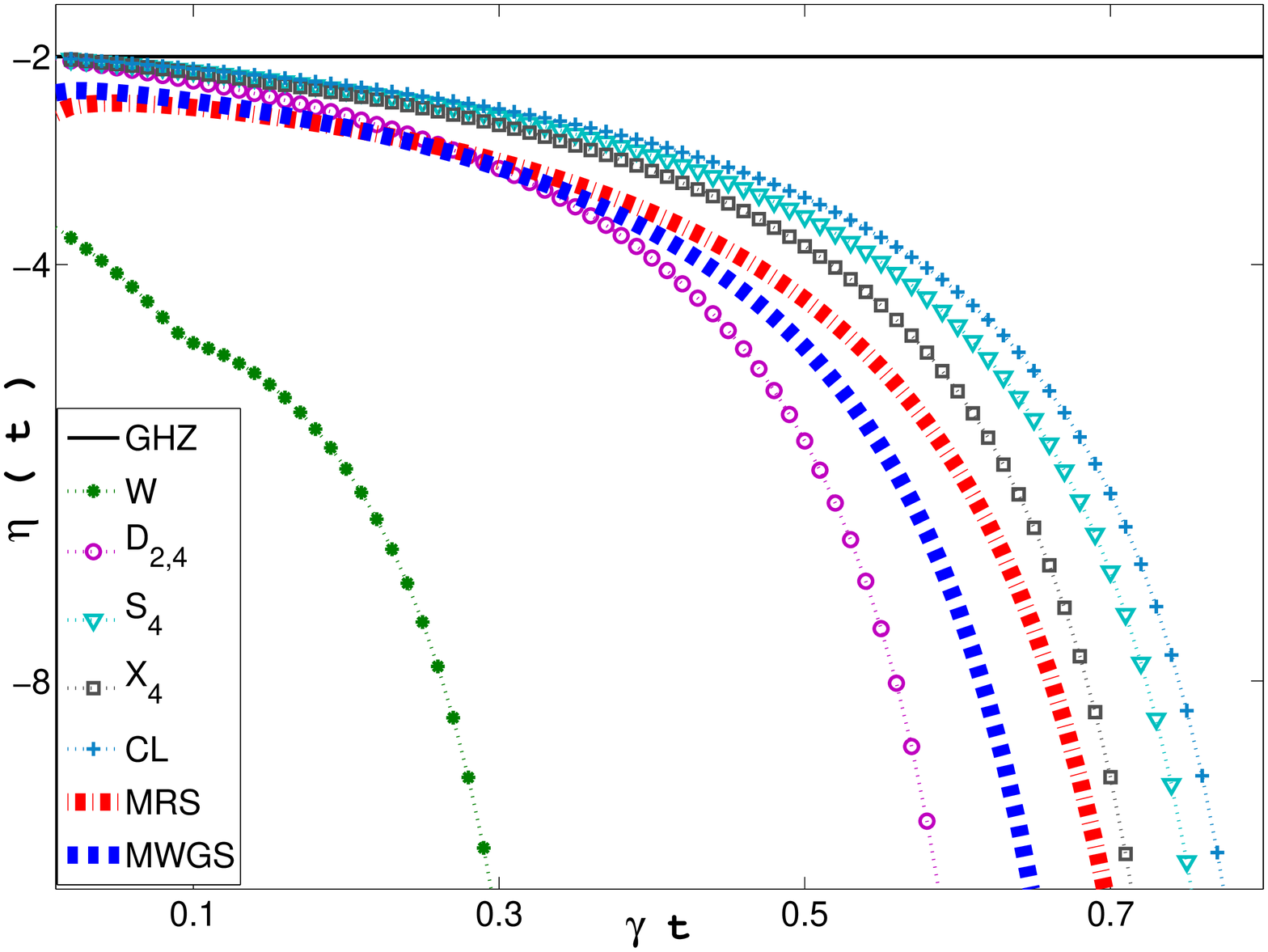}}
\caption{(a) Logarithmic derivative $\eta (t)$ for three qubit
states and under phase damping. MRS and MWGS denote the mean 
values for the  random states, or random weighted graph states,
and CIRS and CIWGS denote the mean values plus/minus the respective 
variances [see Eq.~(\ref{interval})]. 
Here, $|GHZ\rangle$ states  are the most robust states.
(b)~Logarithmic derivative $\eta (t)$ for various four-qubit 
states under phase damping. See text for further details.}
\label{Fig:R3QbPD1}
\end{figure}

For three qubits, we study the robustness of genuine entanglement 
under phase damping for $GHZ$ type states, $W$ type states, random 
pure states and random weighted graph states, 
the results are given in Fig.~\ref{Fig:R3QbPD1}(a). In this figure, 
one may be surprised about the fact that the logarithmic derivative 
for the W state seems to be non-analytic at some point. This, however, has also been
observed for other measures \cite{Guehne-PRA78-2008}. Such points can 
occur, if the entanglement measure is a so-called convex roof measure 
\cite{Horodecki-RMP-2009} and the optimal decomposition in this convex roof construction 
changes qualitatively. In our case, we defined the measure $E(\varrho)$
via a SDP, but the same measure can also be viewed as a convex roof measure \cite{hofmannnew}.

For four qubits, we investigate again the various entangled states mentioned
above, see Fig.~\ref{Fig:R3QbPD1}(b). The variances for the random states 
show the same behavior as before, so we have not displayed them in the 
diagram. 

Summarizing, we can state that for dephasing noise the GHZ state turns 
out to be the most robust state. Other states show a similar behavior
as random states, whereas the $|W\rangle$ state is a very fragile state. 

\subsection{Depolarizing noise}\label{SSec:DN}

\begin{figure}[t!]
\centering
\subfigure[]{\includegraphics[width=7.7cm]{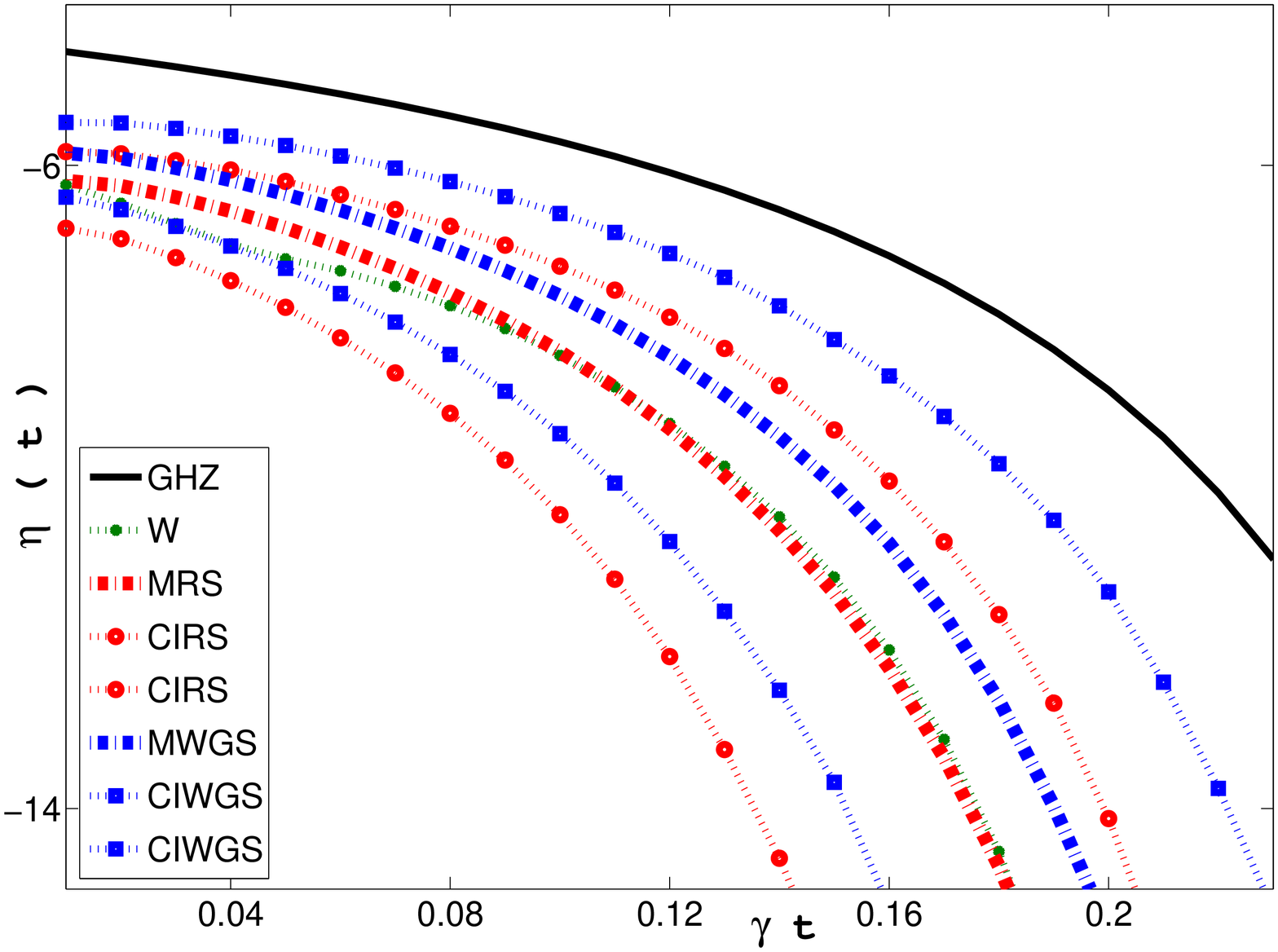}}
\subfigure[]{\includegraphics[width=7.7cm]{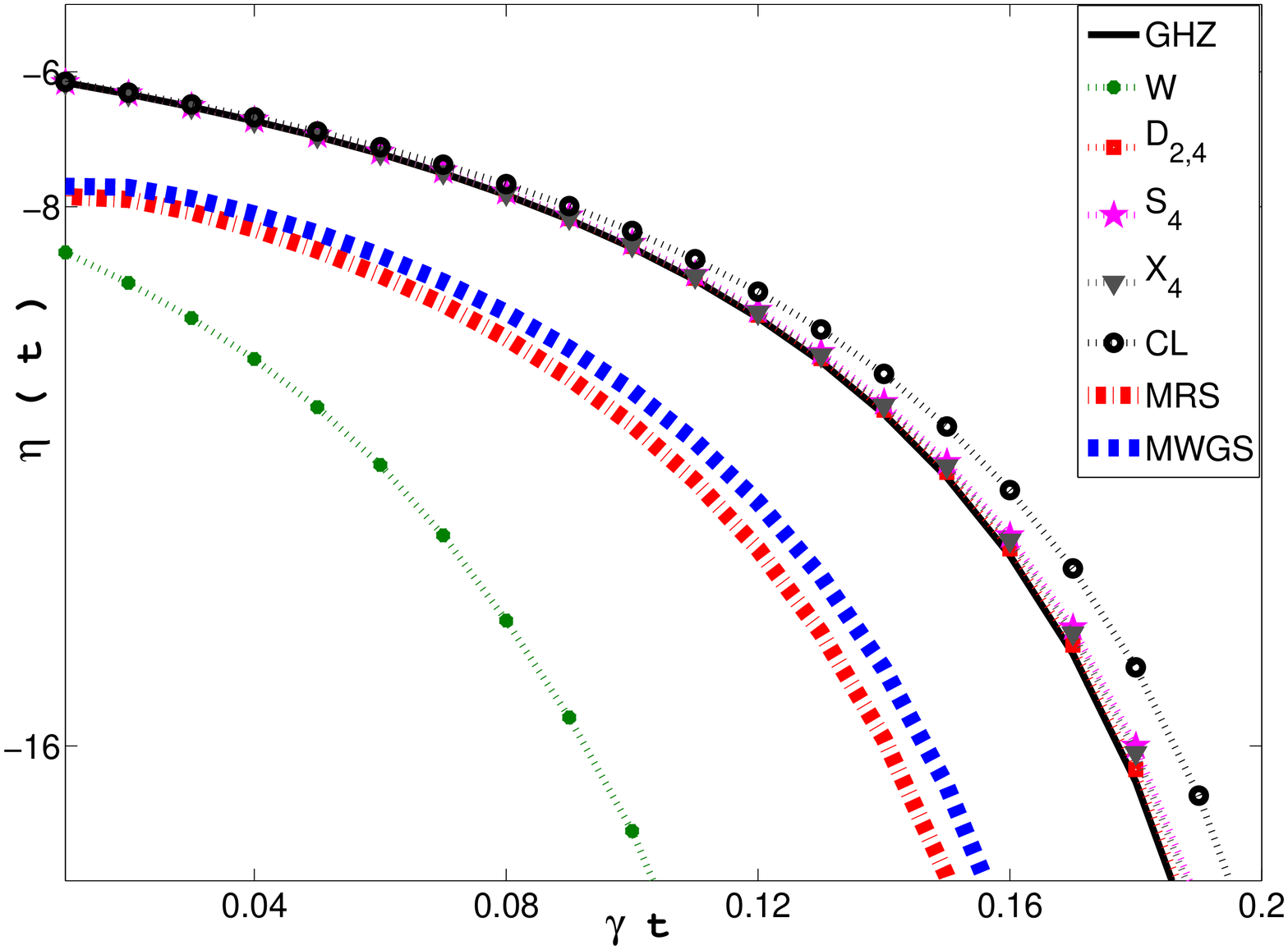}}
\caption{(a) Logarithmic derivative $\eta (t)$ for three qubit
states and under depolarizing noise. MRS and MWGS denote the mean 
values for the  random states, or random weighted graph states,
and CIRS and CIWGS denote the mean values plus/minus the respective 
variances [see Eq.~(\ref{interval})]. 
Here, $|GHZ\rangle$ states  are the most robust states.
(b)~Logarithmic derivative $\eta (t)$ for various four-qubit 
states under depolarizing noise. See text for further details.}
\label{Fig:R3QbDN1}
\end{figure}

Finally, we study the effects of depolarizing noise on 
multiparticle entanglement of three- and four-qubit quantum 
states. For three qubits, the results are given in 
Fig.~\ref{Fig:R3QbDN1}(a). Clearly, the GHZ state is the most 
robust state and the W state does not significantly deviate 
from random states. 

The four-qubit results are given in Fig.~\ref{Fig:R3QbDN1}(b). 
Here, an interesting feature emerges: Apart from the W state, 
all states are significantly better than  random states. From
the states that are better than random states the cluster state
is by a slight amount the most robust state.

\section{Discussion and Summary}
\label{Conc}

We studied the effects of local decoherence on genuine multiparticle
entanglement for various quantum states, including random states. 
We found
that for amplitude damping noise the W-type states are most robust, 
for phase  damping noise the GHZ state is the most robust state, and 
for depolarizing noise several states are significantly more
robust than random states. It is worth mentioning that the robustness
of GHZ and W states has also discussed in Ref.~\cite{lowerbounds}. Our results are
in line with these findings for amplitude damping noise and depolarizing 
noise. For dephasing noise, however, the results of  Ref.~\cite{lowerbounds} are opposite
to ours. A reason for this discrepancy may be, that the lower bound on 
entanglement used in Ref.~\cite{lowerbounds} works better for W class states, 
suggesting that they are more robust to noise. 

There are several directions in which our work can be extended. First, it 
would be very interesting to investigate the scaling of the robustness
with the number of qubits $N$. For that, one needs analytical formulas 
for entanglement measure $E(\vr).$ First results indicate that such 
formulas may be derived \cite{hofmannnew}. Second, it would be interesting
to find  out why certain states are more robust to noise than others. This 
may be investigated by modeling the interaction between the particles and 
the environment. Third, it is of interest to compare for a given multiparticle 
state the robustness of entanglement with the usefulness for some task in 
quantum information processing. For instance, the usefulness for quantum 
metrology can be determined via calculating the Fisher information 
\cite{notall}, so the states which are most robust for metrology can be 
identified. Finally, the robustness of entangled states is also of relevance 
for the characterization of quantum channels \cite{ziman}. Therefore, the 
algorithms used in our paper may be helpful for this task. 

We thank Bastian Jungnitsch, Tobias Moroder, S\"onke Niekamp, Marcel Bergmann 
and Martin Hofmann for discussions. This work has been supported by the EU 
(Marie Curie CIG 293993/ENFOQI) and the BMBF (Chist-Era Project QUASAR). 

\section{Appendix}

\begin{figure}[t!]
\centering
\subfigure[]{\includegraphics[width=7.7cm]{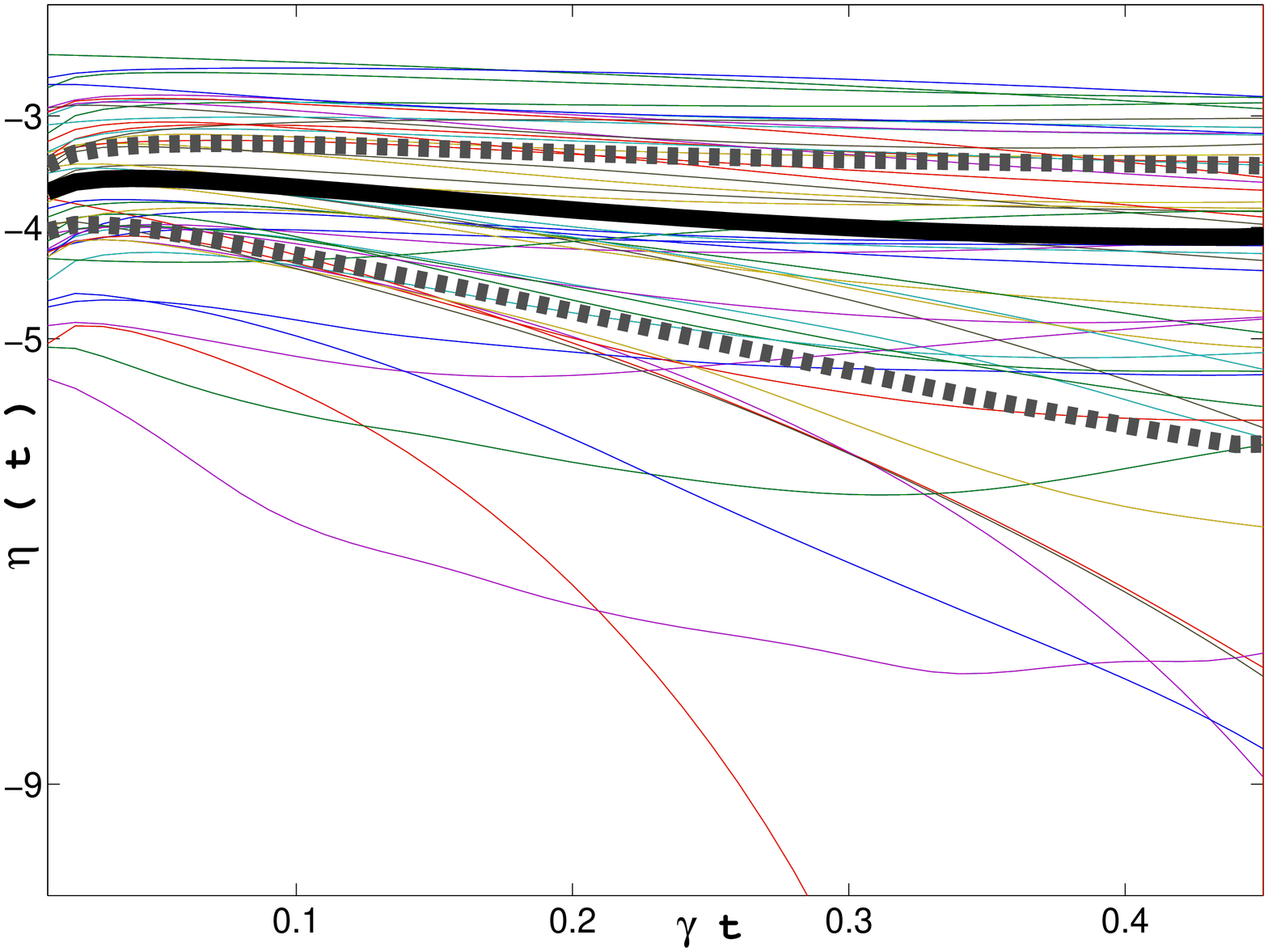}}
\subfigure[]{\includegraphics[width=7.7cm]{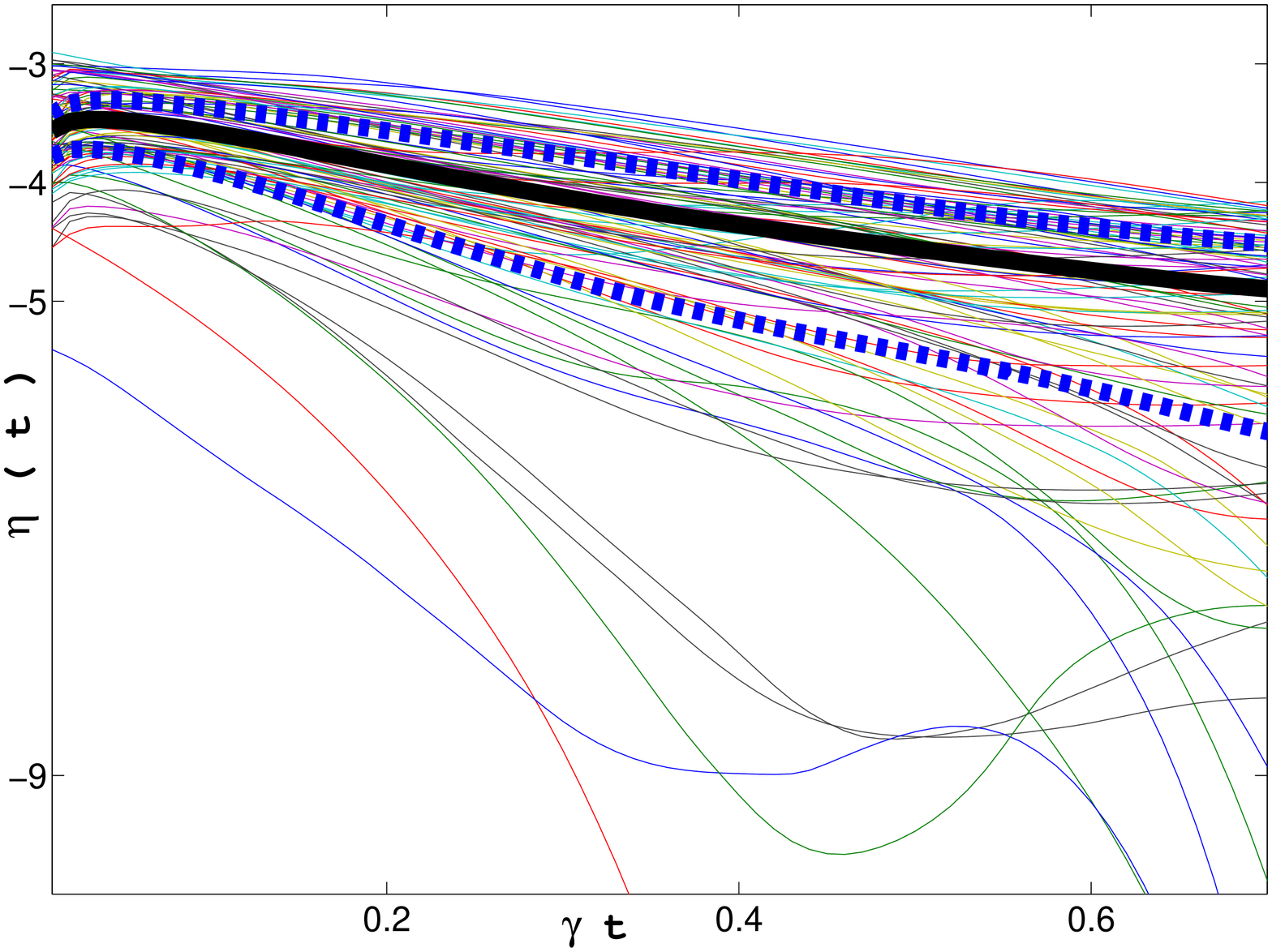}}
\caption{(a) Logarithmic derivative $\eta (t)$ for $100$ random pure states of three qubits under amplitude damping. The MRS is the mean value of $\eta (t)$ for random pure states whereas the thick gray-dashed lines are variances for the same random states (CIRS).
(b) The same as in (a), but for random weighted graph states.}
\label{Fig:rand}
\end{figure}

In order to show the typical behaviour of random states, 
Fig.~\ref{Fig:rand}(a) shows the logarithmic derivative $\eta (t)$
of the entanglement monotone  against parameter $\gamma t$ for $100$ 
random pure states under amplitude damping, while Fig.~\ref{Fig:rand}(b) 
shows the same quantity for random weighted  graph states.

\end{document}